\documentclass[11pt,english]{article}
\usepackage{amssymb}
\usepackage{mathrsfs, amsmath}
\usepackage[dvips]{graphicx}
\usepackage{setspace}
\usepackage{fancyhdr}
\usepackage{xcolor}
\usepackage{ifpdf}
\usepackage{graphicx}
\usepackage{rotating}
\usepackage{comment,tikz}
\usepackage{url}
\definecolor{darkblue}{rgb}{0,0,0.6}
\usepackage[colorlinks=true,citecolor=darkblue,linkcolor=black,urlcolor=darkblue]{hyperref}
\usepackage{cite}
\usetikzlibrary{patterns}
\usepackage[utf8]{inputenc}

\makeatletter
\newcommand*{\defeq}{\mathrel{\rlap{%
                     \raisebox{0.3ex}{$\m@th\cdot$}}%
                     \raisebox{-0.3ex}{$\m@th\cdot$}}%
                     =}
\makeatother

\DeclareMathOperator{\p}{\partial}

\DeclareMathOperator{\Tr}{Tr}

\newcommand{\be}{\begin{equation}}
\newcommand{\ee}{\end{equation}}

\newcommand{\f}{\frac}

\newcommand{\lensp}{S^3/\mathbb{Z}_p}
\newcommand{\lensk}{S^3/\mathbb{Z}_k}
\newcommand{\kp}{\lim_{\substack{k\rightarrow \infty\\p\rightarrow \infty}}}
\newcommand{\kkp}{\lim_{\substack{k/p\rightarrow \infty\\p\rightarrow \infty}}}

\begin{document}
\unitlength = 1mm
\ 

\begin{center}

{ \LARGE {\textsc{\begin{center}Modular invariance on $S^1 \times S^3$ and circle fibrations \end{center}}}}

\vspace{0.8cm}
Edgar Shaghoulian

\vspace{.5cm}

{\it Department of Physics} \\
{\it University of California}\\
{\it Santa Barbara, CA 93106 USA}

\vspace{1.0cm}

\end{center}

\begin{abstract}

\noindent I conjecture a high-temperature/low-temperature duality for conformal field theories defined on circle fibrations like $S^3$ and its lens space family. The duality is an exchange between the thermal circle and the fiber circle in the limit where both are small. The conjecture is motivated by the fact that $\pi_1(S^3/\mathbb{Z}_{p\rightarrow \infty})=\mathbb{Z}=\pi_1(S^1\times S^2)$ and the Gromov-Hausdorff distance between $S^3/\mathbb{Z}_{p\rightarrow \infty}$ and $S^1/\mathbb{Z}_{p\rightarrow \infty} \times S^2$ vanishes.
Several checks of the conjecture are provided: free fields, $\mathcal{N}=1$ theories in four dimensions (which shows that the Di Pietro-Komargodski supersymmetric Cardy formula and its generalizations are given exactly by a supersymmetric Casimir energy), $\mathcal{N}=4$ super Yang-Mills at strong coupling, and the six-dimensional $\mathcal{N}=(2,0)$ theory. For all examples considered, the duality is powerful enough to control the high-temperature asymptotics on the unlensed $S^3$, relating it to the Casimir energy on a highly lensed $S^3$. 
\end{abstract}

\pagebreak
\setcounter{page}{1}
\pagestyle{plain}

\setcounter{tocdepth}{1}

\section{Introduction and conjecture}\label{intro}
The torus is powerful.  It has a nontrivial mapping class group $SL(d,\mathbb{Z})$ which can be used to constrain conformal field theories (CFTs) placed on such a background. As a sample, this symmetry has provided formulas for the asymptotic density of states ($d\geq 2$) \cite{Cardy:1986ie, Shaghoulian:2015kta, Shaghoulian:2015lcn}, monotonicity and sign constraints on the torus vacuum energy ($d>2$) \cite{Shaghoulian:2015kta, Belin:2016yll}, upper bounds on the gaps (above the vacuum) of scaling dimensions and charges of primary operators ($d=2$) \cite{Hellerman:2009bu, Benjamin:2016fhe, Collier:2016cls}, and constraints on operator product expansion coefficients ($d=2$) \cite{Kraus:2016nwo}. 

\begin{figure}
\centering
\includegraphics[scale=0.30]{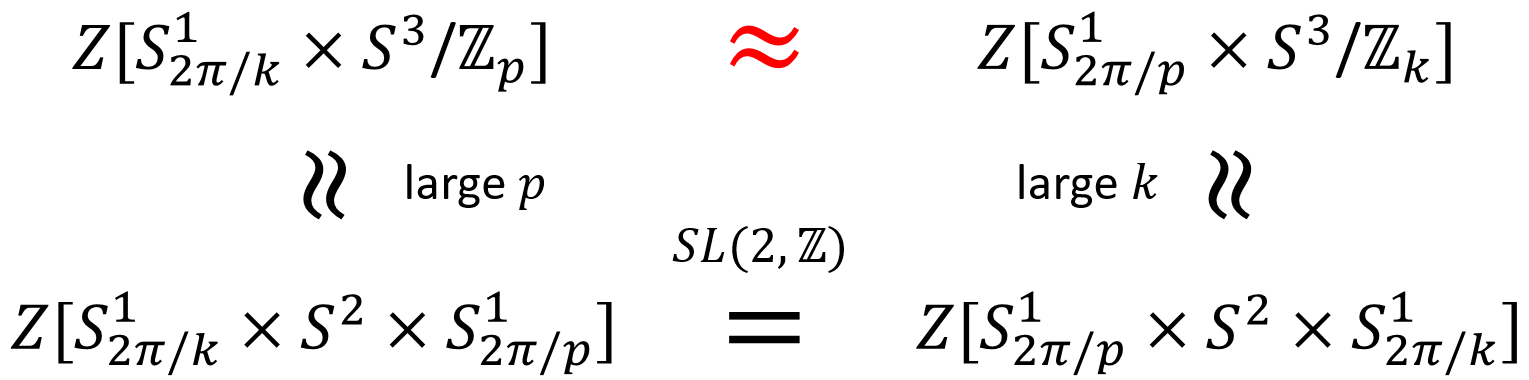}
\caption{\label{boxdiagram} A series of approximate equivalences motivating the conjecture in the case of a lens space partition function. More generally, we can replace $S^3/\mathbb{Z}_k$ with a smooth manifold $\mathcal{M}^{d-1}/\mathbb{Z}_k$. In all cases considered in this paper, the results can be analytically continued to $p=1$ and give a correct equality between the high-temperature partition function on the unlensed manifold and the low-temperature partition function on a highly lensed manifold.} 
\end{figure}

It is also natural to consider CFTs on $S^{d-1}$ due to the state-operator correspondence. In this case there is no general modular property of the theory on $S^1 \times S^{d-1}$ that is known. The special power of a torus is that it is made of circles, and it is the Hamiltonian interpretation that such a circle admits that allows for distinct-looking but fundamentally equivalent quantizations. As it turns out, circles are everywhere.  In this work we will consider circle fibrations $\mathcal{M}^{d-1}$, which are spaces with a free circle action at every point over some base space $\mathcal{M}^{d-2}$. This is often denoted $S^1\longrightarrow \mathcal{M}^{d-1}\longrightarrow \mathcal{M}^{d-2}$, and typical examples include odd-dimensional spheres, e.g. $S^3$ is a circle fibration over $S^2$. Here we argue that a general circle fibration with a freely acting $U(1)$  can be highly quotiented (we will often say ``lensed") into approximating $S^1 \times \mathcal{M}^{d-2}$. There is an emergent circle and the thermal partition function on $S^1/\mathbb{Z}_k \times \mathcal{M}^{d-1}/\mathbb{Z}_p$ inherits an $SL(2,\mathbb{Z})$ invariance when $k$ and $p$ are large with arbitrary ratio. This allows us, for example, to lens the $S^1 \times S^3$ partition function into a $\mathbb{T}^2 \times S^2$ partition function (see figure \ref{boxdiagram}). This then connects the local operator content of the theory -- in a precise and quantifiable way -- to the $\mathbb{T}^2 \times S^2$ partition function, which admits the aforementioned powerful modular symmetry. Interestingly, in all examples considered in this paper the stronger equality 
\be
Z[S^1/\mathbb{Z}_k \times \mathcal{M}^{d-1}] \approx Z[S^1 \times \mathcal{M}^{d-1}/\mathbb{Z}_k]
\ee
 is true, which lets us relate e.g. the high-temperature theory on an unlensed $S^3$ to the low-temperature theory on a highly lensed $S^3$ (here low temperature means $\beta$ large compared to the lensed Hopf fiber of the $S^3$). The rest of this section introduces these ideas in the context of $\mathcal{M}^{d-1}=S^3$.

Given a Lagrangian CFT, one can unambiguously define the partition function on a manifold $\mathcal{M}^{d}$ if the path integral is invariant under large diffeomorphisms (i.e. the mapping class group) of the manifold. This requirement in the case of a two-dimensional CFT on $\mathbb{T}^2=S^1_\beta \times S^1_L$ gives invariance under $SL(2,\mathbb{Z})$, which requires operators to have integer spin and implies a high-temperature/low-temperature duality $Z(\beta/L) = Z(L/\beta)$. For higher-dimensional CFTs on $\mathbb{T}^d$ the theory becomes invariant under $SL(d,\mathbb{Z})$. 
The question of whether high-temperature/low-temperature dualities of this nature extend to CFT partition functions on $S^1_\beta \times S^{d-1}_R$ for $d>2$ has been discussed \cite{Cardy:1991kr} since the case of $d=1$ was understood \cite{Cardy:1986ie}. While free fields have certain properties under modular transformations \cite{Kutasov:2000td}, there is no general behavior.


Much of the power of modular invariance is rather simply stated, and more general: in a Lorentz-invariant theory you are free to label whichever direction you want as time. When doing thermal physics this is an invariance under swapping a thermal cycle with a spatial cycle. Odd-dimensional spheres can be understood as circle bundles, so there is an appetizing $S^1_L$ that one cannot help but try to swap with the thermal $S^1_\beta$. For $\beta = 2\pi/p$ for some non-negative integer $p$, this would mean that the sphere partition function at some temperature would be related to the lens space partition function at some other temperature. This is \emph{not} true in general. However, the conjecture of this paper is that it is true asymptotically as the two cycles shrink to zero size. In other words, we have
\be\label{conj}
\kp Z\left[S^1_{2\pi/k} \times \lensp\right] = \kp Z\left[S^1_{2\pi/p} \times \lensk\right],\qquad p,k\in \mathbb{Z}^+\,.
\ee
The partition function often diverges exponentially in $k$ or $p$ in this limit, so the proper equality is stated in terms of a $k$- or $p$-normalized free energy $\log Z$, or more cleanly by dividing through the equation by e.g. $\kp Z\left[S^1_{2\pi/p} \times \lensk\right]$. I will often leave this implicit. The equality is meant to hold independent of precisely how we take the limit, i.e. for arbitrary ratio $p/k$. 

How can we justify such an equivalence?\footnote{It is tempting to try to derive the modular relation \eqref{conj} and its generalizations by dimensionally reducing along the thermal cycle and fiber cycle to obtain a $(d-2)$-dimensional effective theory. The modular relation would then be the statement that the two different orderings of dimensional reduction commute. However, as we will see through upcoming examples the modular relation connects the leading singular terms in the two quantizations. These are not generically captured by an effective lower-dimensional theory truncated to the zero-mode sector (which instead captures finite pieces), since the leading singularity arises from Kaluza-Klein excitations with wavelengths of the order of the small cycles, which are integrated out in the zero-mode theory. Of course, it should be the case that keeping all the Kaluza-Klein excitations in the dimensionally reduced theories should make them coincide.} In the case of lens spaces, for any finite lensing the two manifolds are not related by a large diffeomorphism. So the natural guess is that the conjectured equivalence will not be true at any finite lensing, which we will see is the case in concrete examples. But the conjecture is meant only to hold in the limit of infinite lensing. Notice that something special is happening in the limit: $\pi_1\left(S^3/\mathbb{Z}_{p\rightarrow \infty}\right) = \mathbb{Z}_{p\rightarrow \infty} = \mathbb{Z}= \pi_1(S^1)$, so it seems that there is an emergent circle justifying the cycle-swapping invariance that would make \eqref{conj} and its generalizations true. The trivial fibration $S^1/\mathbb{Z}_p\times S^2=S^1_{2\pi/p}\times S^2$ and the nontrivial fibration $S^3/\mathbb{Z}_p$ are also approximately isometric for large $p$, a notion which we now make precise.

Consider the Gromov-Hausdorff distance $d_{GH}$, which measures distances between two metric spaces by minimizing over all Hausdorff distances between all possible isometric embeddings of the two metric spaces. In general this is a pseudometric, but since we will be considering compact metric spaces, the distance vanishes if and only if the two spaces are isometric \cite{book:1497252}. In particular this means that the two spaces under consideration are diffeomorphic. Both the lens space $\mathcal{M}_1(p)=S^3/\mathbb{Z}_p$ and $\mathcal{M}_2(p)=S^1_{2\pi/p}\times S^2$ converge to $\mathcal{M}_3=S^2$ as $p\rightarrow \infty$ in the Gromov-Hausdorff sense. This is familiar from dimensional reduction in physics, and is an example of collapse with bounded curvature in mathematics.  This means that $d_{GH}(\mathcal{M}_1(p\rightarrow \infty), \mathcal{M}_3)= 0$ and $d_{GH}(\mathcal{M}_2(p\rightarrow \infty), \mathcal{M}_3)= 0$. But by the triangle inequality we therefore have 
\be
\lim_{p\rightarrow \infty} d_{GH}(S^1_{2\pi/p} \times S^2\,\,,\hspace{2mm} S^3/\mathbb{Z}_p)= 0\,,
\ee
meaning the two spaces become isometric as $p$ becomes large. In particular, for any distance $\epsilon$, we can pick a sufficiently large $p$ such that 
\be
 d_{GH}(S^1_{2\pi/p} \times S^2, S^3/\mathbb{Z}_p)<\epsilon\,.
\ee
Since the spaces become isometric in the limit of infinite $p$, this means we have a  series of approximate isometries in the limit of large $p$ and large $k$ represented in figure \ref{boxdiagram}. 
While an exact isometry between two spaces implies they are homeomorphic, an approximate isometry does not induce any natural notion of ``approximate" homeomorphism. For example we can replace the lens space in figure \ref{boxdiagram} with a squashed (Berger) sphere $S^3_\nu$. This one parameter family of solutions also collapses to an $S^2$ as $\nu\rightarrow 0$, giving the approximate isometries discussed above. However, $\pi_1(S^3_\nu) = 0$ for all $\nu$, meaning it maintains distinct topological properties as $\nu$ is varied and there is not as precise a notion of an ``emergent circle."\footnote{In many examples, however, like CFT$_3$ on $S^3_\nu$, it is still nevertheless true that $Z[S^3_{\nu \rightarrow 0}] \approx Z[S^1_{2\pi \nu} \times S^2]$.}

These arguments suggest that the lens space partition function at high temperature and large lensing inherits the mapping class group of the $\mathbb{T}^2\times S^2$ partition function. This justifies the cycle-swapping equivalence conjectured in \eqref{conj}. (But notice that it does not provide a justification for the stronger equality \eqref{strongereq}.)

The conjecture of this work can be stated more generally as an invariance under swapping the thermal cycle with the cycle of any manifold  $\mathcal{M}^{d-1}$ that can be written as a circle bundle with a (possibly locally) freely acting $U(1)$, which allows lensing the cycle in $\mathcal{M}^{d-1}$ to become arbitrarily small and induce collapse to $\mathcal{M}^{d-2}$. The size of the emergent $S^1$ is set by Vol$(S^1/\mathbb{Z}_p \times \mathcal{M}^{d-2})$ = Vol$(\mathcal{M}^{d-1}/\mathbb{Z}_p)$. Altogether, the requirement for replacing $\mathcal{M}^{d-1}/\mathbb{Z}_p$ with $S^1/\mathbb{Z}_p \times \mathcal{M}^{d-2}$ will be that the two spaces become isometric and have the same fundamental group as the lensing is taken large.

A subtle point about this equivalence is that it will imply that the high-temperature partition function is equal to a low-temperature partition function. The high-temperature partition function defines a scheme-independent thermal entropy, while the low-temperature partition function depends on the vacuum energy. As is well known, the vacuum energy on curved manifolds is generically scheme-dependent in even-dimensional CFT. What gives? The resolution to this is that the vacuum energy becomes scheme-independent as the fiber cycle is taken small. In this case, the contribution of any scheme-dependent counterterm goes to zero. This is general and independent of supersymmetry. It is explained more carefully in the case of the trivial fibration $\mathbb{T}^2\times S^2$ in the next section. 

Taking the order of a quotient to be infinitely large as a control parameter does not seem to have received much (if any) attention in the literature. This note will show that it provides nontrivial constraints at zeroeth order in perturbation theory around the infinite quotient, with the potential to go to higher orders. For a distinct application of infinite quotients see \cite{Benini:2011nc}.

Related work includes \cite{Duff:1998cr, Hosseini:2016cyf, Brunner:2016nyk, Zhou:2015cpa}.

\subsection{$\mathbb{T}^2\times S^2$ and scheme independence of vacuum energy}
Consider a CFT on $\mathbb{T}^2\times S^2$. To match the notation of the eventual lens space calculations, we will consider cycles of lengths $\beta=2\pi/k$ and $\beta'=2\pi/p$ at large $p$ and $k$, although in this case taking large $p$ and $k$ is not necessary to obtain an invariance under $p\leftrightarrow k$. We have
\be
\kp Z(2\pi/k)\equiv \kp Z\left[S^1_{2\pi/k} \times S^1_{2\pi/p}\times S^2\right] =\kp Z\left[S^1_{2\pi/p} \times S^1_{2\pi/k}\times S^2\right] \equiv \kp Z(2\pi/p) \,.
\ee
The argument of $Z(\dots)$ denotes the size of the thermal cycle. The argument of $Z[\dots]$ denotes the entire manifold, where the first factor represents the thermal cycle. On the left-hand-side we have an inverse temperature $\beta=2\pi/k$ and spatial volume $V=4\pi \beta'=8\pi^2/p$, while on the right-hand-side we have an inverse temperature $\beta'=2\pi/p$ and spatial volume $V'=4\pi \beta = 8\pi^2/k$. This means in the limit of large $k/p$ we have $\beta/V \ll 1$, so the left-hand-side represents a high-temperature partition function, while $\beta'/V'\gg 1$, so the right-hand-side represents a low-temperature partition function. This high-temperature/low-temperature duality implies that the vacuum energy governs the asymptotic entropy:
\be
\beta\rightarrow 0: \qquad S(\beta) = (1-\beta \p_\beta)\log Z(\beta) =  (1-\beta \p_\beta)\log Z(\beta')=(1-\beta \p_\beta)\left(-\beta' E_{\textrm{vac},\, S^1_{\beta}\times S^2}\right).
\ee
The high-temperature thermal entropy is scheme-independent, while the vacuum energy is generically scheme-dependent. In the case of two-dimensional CFT and the Cardy formula, the vacuum energy is scheme-independent and the formula that relates it to the thermal entropy is therefore sensible. In higher dimensions, the vacuum energy is scheme-independent if the spatial background has no curvature, which is why the higher-dimensional Cardy formula of \cite{Shaghoulian:2015lcn, Shaghoulian:2015kta} is also sensible. But now we are on a curved manifold $\mathbb{T}^2 \times S^2$. There is a dimensionless counterterm, $b\int d^4 x \sqrt{g} R^2$ with $b$ arbitrary, which we can write in the action and which will shift the answer for the vacuum energy by an amount proportional to $b$ (see \cite{Assel:2015nca} for a nice discussion). How can the above formula be consistent given this ambiguity? The answer is simple: in the limit we are considering the vacuum energy is on $S^1_{\beta'} \times S^2$ with $\beta'\rightarrow 0$. The shift ambiguity due to the contribution of the counterterm is proportional to $\beta'$ and vanishes in this limit, rendering the vacuum energy physical. This explanation is general and controls the scheme-independence of the vacuum energy in all upcoming examples.


\section{Four dimensions}\label{checks}
\subsection{Conformally coupled scalar}
Consider a conformally coupled scalar at inverse temperature $\beta=2\pi/k$ on a lens space $S^3_R/\mathbb{Z}_p$ with radius $R$. This space has volume $V=2\pi^2R^3/p$. At high temperature $\beta^3/V\ll 1$ we have\footnote{I will almost always write expressions that are formally divergent. For example, for \eqref{ex} the precise thing to do is to define a density by taking the limit of $pk^{-3}\log Z$, which would give a finite answer.}
\be
\kkp\log Z\left[S^1_{2\pi/k} \times S^3_R/\mathbb{Z}_p\right] = \f{\pi^4 R^3}{45p(2\pi/k)^3}=\f{\pi k^3R^3}{360p}\label{ex}
\ee
Swapping the thermal cycle with the fiber cycle gives us a low-temperature partition function on a different lens space, for which we have
\be
\kkp\log Z\left[S^1_{2\pi R/p} \times S^3_R/\mathbb{Z}_{kR}\right] = -\left(\f{2\pi R}{p}\right)\left(-\f{k^3R^3}{720R}\right)=\f{\pi k^3R^3}{360p}\,.
\ee
The above expression illustrates that the low-temperature partition function projects to the vacuum state with energy that becomes scheme-independent as $k\rightarrow \infty$:
\be
E_{\textrm{vac},\,S^3_R/\mathbb{Z}_{kR}}=\f{14-10(kR)^2-(kR)^4}{720R(kR)}\longrightarrow \f{-k^3R^3}{720R}
\ee
The expressions above are straightforward to calculate and can be found for example in \cite{Asorey:2012vp}. We see that the two expressions are equal as predicted by the conjecture. For the rest of the checks, we will not keep the radius of the lens space explicit.

\subsection{$\mathcal{N}=1$ superconformal theories and Di Pietro-Komargodski formula}\label{zohar}
In \cite{DiPietro:2014bca}, asymptotic Cardy-like formulas for supersymmetric partition functions in four and six dimensions were proposed (see also \cite{Ardehali:2014zba, Ardehali:2014esa} for early conjectures in this direction). Here we will focus on the case of $\mathcal{N}=1$ superconformal theories in four dimensions. Supersymmetry will be preserved so the ``thermal" $S^1$ will have non-thermal periodicity conditions that match those of the Hopf fiber.

We first consider the case of a squashed lens space, a generalization of \cite{DiPietro:2014bca} treated in \cite{DiPietro:2016ond}. In the literature both the Berger sphere and an ellipsoid are often referred to as squashed spheres. Here we consider the Berger sphere, and we will later consider the ellipsoid. The Berger sphere metric can be written as
\be\label{hopf}
ds^2=\f{1}{4}\left(d\theta^2+\sin^2\theta\, d\phi^2+\nu^2\left(d\psi+\cos\theta \,d\phi\right)^2\right)
\ee
where $\psi \sim \psi+4\pi$. The $\nu=1$ point is the unit three-sphere. The high-temperature supersymmetric partition function is given as 
\be\label{sq1}
\kkp \log Z\left[S^1_{2\pi\nu/k} \times S^3_\nu/\mathbb{Z}_p\right]=-\f{8\pi k}{3p}(a-c)\,.
\ee
The thermal circle is normalized to match the size of the emergent circle in the lens space when $k=p$. The size of the emergent circle is fixed by the volume condition discussed in the introduction, i.e. equating Vol$[S^3/\mathbb{Z}_p]=$Vol$[S^1 \times S^2]$ for $S^2$ with radius $1/2$ sets the size of the emergent $S^1$ to be $2\pi \nu/p$. Since the Hopf fiber size is fixed, it is easy to see that the emergent circle must have size $2\pi\nu/p$. The supersymmetric vacuum energy on a lensed Berger sphere $S^3_\nu/\mathbb{Z}_k$ is given as \cite{Martelli:2015kuk}
\be\label{susyberger}
E_{\text{susy, } S^3_\nu/\mathbb{Z}_k} = \f{16}{27k\nu}(3c-2a)+\f{4k}{3\nu}(a-c)\,.
\ee
Unlike the previous section, this vacuum energy is manifestly scheme-independent for arbitrary fiber size, thanks to supersymmetry. The low-temperature partition function is given by the supersymmetric vacuum energy through a projection to the ground state: 
\be\label{sq2}
\kkp \log Z\left[S^1_{2\pi\nu/p} \times S^3_\nu/\mathbb{Z}_k\right]=-\left(\f{2\pi\nu}{p}\right)\left(\f{4k}{3\nu}\right)(a-c)\,.
\ee
The cited works consider a sphere of radius $2$, so a rescaling of those results was done in \eqref{sq1}-\eqref{sq2}. As advertised, the high-temperature partition function on a given squashed lens space equals the low-temperature partition function on a different squashed lens space (the amount of squashing remains the same but the amount of lensing is different). 

It is somewhat surprising that the equivalence holds at $\mathcal{O}(k)$ for supersymmetric partition functions, since for non-supersymmetric partition functions -- which have an extensive leading term $\mathcal{O}(k^3)$ -- we will see in upcoming sections that the equivalence fails at $\mathcal{O}(k)$ (we did not calculate the conformally coupled scalar to this order to see disagreement). Thus, agreement at $\mathcal{O}(k)$ seems special to supersymmetric partition functions. 

We also see from the above results that the large squashing limit $\nu \rightarrow 0$ (where we do not put a factor of $\nu$ in the size of the thermal circle now), which also collapses the manifold to $S^2$, does not give an appropriate modular equivalence, since it does not isolate the second term in \eqref{susyberger}. In particular, $Z[S^1_{2\pi/k} \times S^3_\nu] \neq Z[S^1_{2\pi \nu} \times S^3_{k^{-1}}]$ at large $k$ and small $\nu$, except for the trivial case $\nu = k^{-1}$. Presumably the approach to equivalence of fundamental groups as the manifold collapses is important for supersymmetric free energies but not as important for extensive free energies.


Now consider the theory on an ellipsoidal lens space $S_b^3/\mathbb{Z}_p$, with $b=1$ giving the round three-sphere lens space. The metric for the ellipsoid can be written in toroidal coordinates as
\be
ds^2=(b^2 \cos^2\theta + b^{-2}\sin^2\theta)d\theta^2 + b^2 \sin^2\theta\, d\varphi_1^2+b^{-2}\cos^2\theta d\varphi_2^2\,.
\ee
where $\theta\in[0,\pi/2]$ and $\varphi_{1,2}\sim \varphi_{1,2}+2\pi$.   Lensing performs the identification $(\varphi_1, \varphi_2)=(\varphi_1+2\pi/p, \varphi_2-2\pi/p)$. The Hopf coordinates are given by  $\psi=b\varphi_1+b^{-1}\varphi_2$, $\chi=b\varphi_1-b^{-1}\varphi_2$. The $U(1)$ isometry of the $\psi$ coordinate is that of the Hopf fibration and acts freely. However, for general irrational squashing $b$ the orbits of $\partial_\psi$ do not close except at the poles $\theta=0$, $\theta=\pi/2$ (see \cite{Tanaka:2012nr, Hosomichi:2014hja, Willett:2016adv} for a discussion of this and its relevance in computing supersymmetric vortex/Wilson loops \cite{Drukker:2012sr, Kapustin:2012iw}). The volume condition enters the stage and fixes the size of the emergent circle, which is found to be $2\pi/p$. The simplest way to see this is to consider the unlensed ellipsoid with rational $b^2=r/s$ for coprime $r,\,s$, where the closed loops in $\theta\in (0,\pi/2)$ form an $(r,s)$-torus knot of length $2\pi$ \cite{Tanaka:2012nr}. 

The high-temperature partition function is given as \cite{DiPietro:2014bca}
\be
\kkp \log Z\left[S^1_{2\pi/k} \times S^3_b/\mathbb{Z}_p\right]=-\f{4\pi k(b+b^{-1}) }{3p}\,(a-c)\,.
\ee
The supersymmetric vacuum energy \cite{Cassani:2014zwa, Assel:2014paa} on a lensed ellipsoid is given as \cite{Assel:2015nca, Martelli:2015kuk}
\be\label{cas}
E_{\textrm{susy}, \,S^3_b/\mathbb{Z}_k}=\f{2 k(b+b^{-1}) }{3}\,(a-c)+\f{2}{27k}(b+b^{-1})^3(3c-2a)\,,
\ee
so the low-temperature partition function is given as
\be
\kkp \log Z\left[S^1_{2\pi/p} \times S^3_b/\mathbb{Z}_k\right] = \kkp -\left(\f{2\pi}{p}\right)\left(\f{2 k(b+b^{-1}) }{3}\,(a-c)\right)
\ee
The two expressions are equal as required. 

Notice that the duality is powerful enough to control the high-temperature partition function on an unlensed $S^3_{b,\nu}$ (which in the case of a round $S^3$ is counting local operators that sit in short representations of the superconformal group), equating it with the low-temperature partition function on a highly lensed $S^3_{b,\nu}$. Also notice that the superconformal index, which is shifted from the supersymmetric partition function by a vacuum energy factor $e^{\beta E_{\text{susy}}}$, would not exhibit nice modular properties. This precisely mimics the case of modular invariance in two-dimensional CFTs, where the shift of operator dimensions by $-c/12$ is necessary to exhibit modular invariance.

\subsection{$\mathcal{N}=4$ super Yang-Mills at weak coupling}
Now consider $\mathcal{N}=4$ $SU(N)$ super Yang-Mills theory at weak 't Hooft coupling. The relevant partition functions and vacuum energies we will use in this section have been calculated in \cite{Hikida:2006qb}. 

First we consider the NS-NS partition function, which corresponds to antiperiodic periodicity conditions for the fermions along the thermal cycle and the fiber cycle (requiring $k$ to be even). We have
\begin{align}
\kkp\log Z\left[S^1_{2\pi/k} \times \lensp\right] &= \f{\pi N^2k^3}{24p}\,,\\
 \kkp\log Z\left[S^1_{2\pi /p} \times \lensk\right]&= -\f{2\pi}{p}\,E^{(\text{NS})}_{\textrm{vac},\,\lensk} =\f{\pi N^2k^3}{24p}\,,
\end{align}
where we have used the NS vacuum energy 
\be
E^{(\text{NS})}_{\textrm{vac},\,\lensk} = N^2\left(\f{3}{16k}+\f{k}{12}-\f{k^3}{48}\right).
\ee

Now consider NS-R boundary conditions, which corresponds to picking the fermions to be antiperiodic along the thermal cycle and periodic along the fiber cycle. Then the conjecture claims that the thermal partition function for the theory with periodic fermions along the fiber cycle will be related to the twisted partition function for the theory with antiperiodic fermions along the fiber cycle:
\begin{align}
 \kkp \Tr \left[\exp\left(-\f{2\pi}{k}\, H_{\lensp}\right)\right] =\kkp \Tr\left[ (-1)^F \exp\left(-\f{2\pi}{p}\, H_{\lensk}\right)\right] .
\end{align}
The high-temperature NS-R partition function can be calculated and is the same as the high-temperature NS-NS partition function from before. The low-temperature R-NS partition function projects to the bosonic vacuum, so also gives the the same result as the NS-NS partition function from before. The equality therefore works out in this case as well. The choice of non-thermal periodicity conditions does not spoil anything since we maintained an extensive free energy.


\subsection{$\mathcal{N}=4$ super Yang-Mills at strong coupling}
We can also check the conjecture at strong coupling using the AdS/CFT correspondence. In fact, for this case we will be able to illustrate modular $S$-invariance at intermediate temperatures since we will have access to log $Z$ as a function of arbitrary $k$ and $p$. This requires using the proposed phase structure on $S^1_{2\pi/k} \times S^3/\mathbb{Z}_p$, with antiperiodic fermions along both cycles, for which there is strong evidence but no proof. The bulk phase structure has two saddles, the thermal Eguchi-Hanson-AdS (EH) metric representing the confined phase and the AdS-Schwarzschild$/\mathbb{Z}_p$ black hole (BH) representing the deconfined phase. Both solutions can be given as 
\be
ds^2= -f(r)dt^2+\f{dr^2}{f(r)h(r)}+\f{r^2}{4}\left(d\theta^2+\sin^2\theta\,d\phi^2+h(r)\left(d\psi+\cos \theta\, d\phi\right)^2\right),
\ee
where the functions are given as
\begin{align}
\text{AdS-Schwarzschild$/\mathbb{Z}_p$ black hole}: \quad f(r) &= 1-\f{\mu}{r^2}+\f{r^2}{\ell^2}\,, \hspace{10mm}h(r)=1\,,\\
\text{Eguchi-Hanson-AdS (ground state)}: \quad f(r) &= 1+\f{r^2}{\ell^2}\,,\hspace{19mm}h(r)=1-\f{\ell^4(p^2/4-1)^2}{r^4}\,,
\end{align}
for the two solutions. The lensing is due to the periodicity $\psi \sim \psi+4\pi/p$. The on-shell actions are then given as
\begin{align}
 \log Z_{\text{BH}}= \frac{\pi ^2 k \left(\left(\sqrt{1-\frac{8}{k^2}}+1\right) k^2-8 \sqrt{1-\frac{8}{k^2}}-12\right)}{128 G p}\,,\qquad
\log Z_{\text{EH}} = \frac{\pi ^2 \left(p^4-8 p^2+4\right)}{64 G k p}\,.
\end{align}
There is a phase transition between these two saddles at
\be
k_\star= \frac{p^4-8 p^2+4}{\sqrt{-6 p^4+48 p^2-88+\left(p^4-8 p^2+20\right)^{3/2}}}\,.
\ee
We can test the proposed modular invariance at medium temperatures due to the exact expressions above. At large $k$ and $p$, independent of how we take the limit (i.e. independent of exactly what ratio of powers of $k$ to $p$ we keep fixed), we have 
\be
k\gg 1, \,\,p\gg 1:\quad \log Z_{\text{BH}}=\f{\pi ^2 k^3}{64 p G}-\f{3 \pi ^2 k}{16 pG}+\mathcal{O}\left(\f{1}{kp}\right)\,,\qquad \log Z_{\text{EH}}=\f{\pi ^2 p^3}{64 k G}-\f{\pi ^2 p}{8 kG}+\mathcal{O}\left(\f{1}{kp}\right)\,.
\ee
As advertised, the leading order answer is invariant under swapping $k\leftrightarrow p$, since this maps us from the confined (deconfined) to the deconfined (confined) phase, leaving the partition function invariant. Notice that the first subleading correction ruins this invariance, illustrating the necessity of the limit. As required by modularity, the phase transition at large $p$ occurs at $k_\star=p$. This is just like the BTZ/thermal AdS$_3$ transition, which by modularity necessarily occurs when the thermal cycle size equals the spatial cycle size ($\beta = L$). 

As in the previous section, we can also test whether the NS-R partition function equals the R-NS partition function under a swap $k\leftrightarrow p$. The NS-R theory does not admit the Eguchi-Hanson-AdS metric, so AdS-Schwarzschild$/\mathbb{Z}_p$ dominates the ensemble for all $k/p$. The R-NS theory does not admit AdS-Schwarzschild$/\mathbb{Z}_p$, so the Eguchi-Hanson-AdS metric dominates the ensemble for all  $k/p$. The two saddles map to one another under the cycle swap, and the equivalence of the NS-R and R-NS partition function under a swap $k\leftrightarrow p$ at large $k$ and $p$ comes out correct. Notice that by by considering both $k/p>1$ and $k/p<1$ this covers the two possible spin structures NS-R and R-NS.

\section{Six dimensions}
\subsection{$\mathcal{N}=(2,0)$ theory on circle fibration over $S^2 \times S^2$}
We can test the conjecture for the strongly coupled $\mathcal{N}=(2,0)$ theory. Here we consider supergravity in AdS$_7$ with asymptotic topology $S^1 \times \mathcal{M}^5/\mathbb{Z}_p$. We will consider $\mathcal{M}^5$ to be a circle fibration over $S^2 \times S^2$. We again have two competing saddles, a black hole solution representing the deconfined phase and a solitonic solution representing the confined phase. The relevant solutions and the transition between the two saddles were developed in \cite{Awad:2000gg, Mann:2003zh, Stotyn:2009ff}.

Both solutions can be given as
\be
ds^2=-f(r)dt^2+\f{dr^2}{f(r)h(r)}+\f{r^2}{6}\left(\sum_{i=1}^2 d\phi_i^2+\sin^2\theta_i d\phi_i^2\right)+\f{r^2}{9}\,h(r)\left(d\psi+\sum_{i=1}^2 \cos\theta_i d\phi_i\right)^2\,,
\ee
where the functions are given as 
\begin{align}
\text{Black hole}: \qquad f(r) &= 1-\f{\mu}{r^4}+\f{r^2}{\ell^2}\,, \hspace{20mm}h(r)=1\,,\\
\text{Ground state}: \qquad f(r) &= 1+\f{r^2}{\ell^2}\,,\hspace{29mm}h(r)=1-\f{(p^2/4-1)^3}{r^6}\,,
\end{align}
The coordinate $\psi$ has periodicity $4\pi/p$, but the relevant  \emph{proper} periodicity is $4\pi/(3p)$, so we normalize the thermal cycle to have periodicity $4\pi/(3k)$. This means the two relevant circles will have the same periodicity when $p=k$. The on-shell actions are then given as
\begin{align}
 \log Z_{\text{BH}}&=\f{\pi ^3 \left(\left(\sqrt{9 k^2-96}+3 k\right)^6/373248-\left(\sqrt{9 k^2-96}+3 k\right)^4/2592+5\right)}{162 k p}\,,\\
\log Z_{\text{EH}} &= \f{\pi ^3 \left(8 \left(p^2/4-1\right)^3+5\right)}{162 k p}\,.
\end{align}
The black hole expression corrects a mistake in eqn. (17) of \cite{Stotyn:2009ff}. Taking the large $p$ and large $k$ limit, again in any order, gives us
\be
k\gg 1, \,\,p\gg 1:\quad \log Z_{\text{BH}}=\f{\pi ^3 k^5}{1296 p}-\f{5 \pi ^3 k^3}{324 p}+\mathcal{O}(k)\,,\qquad \log Z_{\text{EH}}=\f{\pi ^3 p^5}{1296 k}-\f{\pi ^3 p^3}{108 k}+\mathcal{O}(p)\,.
\ee
This has the proposed $k\leftrightarrow p$ invariance at leading order, and we again see that it is violated at first subleading order. The phase transition between saddles occurs at $p=k$ as required by modularity. This context and that of $\mathcal{N}=4$ in the last section have phase structures that can be predicted (at large $k$ and $p$) using modular invariance and center symmetry arguments as explained in \cite{Shaghoulian:2016xbx}.

\subsection{$\mathcal{N}=(1,0)$ theories on $S^5$}
Di Pietro and Komargodski also propose a Cardy-like formula in six dimensions. Checking that this agrees with the conjecture of this paper is left for future work. A more basic check, however, is satisfied: the squashing parameters $\omega_1$, $\omega_2$, and $\omega_3$ of the five-sphere \cite{Imamura:2012bm}, which enter at leading order in high temperature in the Cardy-like formula as $\log Z \sim 1/(\omega_1\omega_2\omega_3)$, enter into the supersymmetric vacuum energy in the same way \cite{Bobev:2015kza, Bak:2016vpi}.

\section{Outlook}\label{conclusions}
In the context of odd-dimensional spheres, the conjecture of this work relates sphere partition functions to torus partition functions. This allows the use of modular invariance while maintaining an interpretation in terms of the operator content of the theory.

Only modular $S$-invariance has been checked in this paper, for a handful of conformal field theories, on a few different circle bundles (ellipsoids and Berger spheres in four dimensions and circle bundles over $S^2 \times S^2$ in six dimensions). The most general form of the conjecture is that for any circle fibration with a locally free $U(1)$ action that admits arbitrarily large lensing of the fiber (e.g. Seifert manifolds in three dimensions\footnote{The simplest case of Seifert manifolds are circle bundles over a Riemann surface, for which there is a globally free $U(1)$ admitting arbitrary lensing by $\mathbb{Z}_p$. The case of circle bundles over orbifolds (which still give a smooth three-dimensional geometry) is a little different. The orbifold is characterized by $n$ integers  $\alpha_i$ which denote the conical deficit at various points on the orbifold \cite{FURUTA199238} (see \cite{Beasley:2005vf} for a nice exposition). In this case quotients by $\mathbb{Z}_{\alpha_i}$ have fixed points. By only considering quotients $\mathbb{Z}_p$ with $p$ relatively prime to $\alpha_i$, we have a smooth family of arbitrarily large quotients.}), there will emerge an $SL(2,\mathbb{Z})$ invariance of the theory in terms of the complex structure $\tau$ of the emergent $\mathbb{T}^2$. Phrasing the invariance in terms of the complex structure requires rigid rescalings of the base manifold.
It would be interesting to test this general $SL(2,\mathbb{Z})$ invariance and study any  modular covariance properties of correlators under the symmetry.


The cycle swapping analyzed in this paper does not require conformality. As far as the same procedure on $\mathbb{T}^2$ is concerned, it can be understood in terms of Euclidean rotational invariance. So it should be true for generic quantum field theories. It would be interesting if these ideas shed light on the 2d modularity of 4d large-$N$ confined theories discussed in \cite{Basar:2015xda, Basar:2015asd}.

It is necessary for the vacuum energy to become scheme-independent to enter into a Cardy-like formula relating it to a thermal entropy. This is guaranteed by a shrinking circle fiber. Interestingly, supersymmetric theories admit a scheme-independent vacuum energy without shrinking circle fiber. This makes it possible (and if you are an optimist like myself, it seems to suggest) that in supersymmetric theories the vacuum energy enters into Cardy-like formulas beyond the shrinking fiber limit. This would mean a high-temperature/low-temperature duality beyond the large $k$ and large $p$ limit necessary in general.

There are three important extensions to this work that would be nice to map out. The first is to build a ``$1/k$ expansion" to the leading-order modular invariance. As we saw, at subleading order in $1/k$ the equality breaks down as expected. Nonperturbatively, the breakdown for holographic theories is $\mathcal{O}(1)$: the Hawking-Page transition on $S^1 \times S^{d-1}$ occurs at $\beta=2\pi/(d-1)$, while a cycle-swapping equivalence would predict $\beta = 2\pi$. The simplicity of the disagreement for $d>2$ begs for an explanation.  

The second extension concerns quotients that are not freely acting, i.e. orbifolds. This is important in two applications. First, for odd-dimensional CFTs the only finite, freely acting quotients of $S^{2n}$  ($n\in \mathbb{Z}^+$) are by $\mathbb{Z}_2$. However, the mathematics and some of the physics of orbifolds is understood. For example, the orbifold version of Seifert-van Kampen's theorem implies $\pi_1(S^{2n}/\mathbb{Z}_p) = \mathbb{Z}_p$. Physical quantities are usually calculated by regulating the conical metrics into smooth metrics and taking a limit at the end, or (equivalently) by insertions of delta function sources in the curvature at the orbifold points \cite{Fursaev:1995ef, Banados:1993qp}. As long as a complete and consistent set of regulator-independent rules exist for this scenario, then we can e.g. lens the two-sphere into a two-football and induce collapse. The second application of this technology is to $S^3$ represented as a torus fibration over the interval, where now we can lens both $S^1$s of the torus fibration, instead of just the Hopf fiber created by a linear combination of the two. This generically gives orbifold geometries and may lead to an emergent $SL(3,\mathbb{Z})$ symmetry of the thermal partition function.

The third extension would be to study modular covariance of correlation functions. The natural expectation is that at leading order all correlation functions will inherit the modular covariance properties of the emergent $\mathbb{T}^2$.

\subsection{Unlensed $S^1 \times S^3$ partition function}
In every case considered in this work, the conjectured modular invariance turns out to control the high-temperature limit $\beta = 2\pi/k \rightarrow 0$ on an unlensed $S^3$, relating it to the low-temperature partition function on $S^3/\mathbb{Z}_{k\rightarrow \infty}$. For thermal partition functions of local CFTs this is not too surprising, as the high-temperature partition function is rather universal and depends only on the volume of the spatial manifold. To be consistent with the formulas of \cite{Shaghoulian:2015lcn, Shaghoulian:2015kta}, this implies that the coefficient controlling the vacuum energy on $S^3/\mathbb{Z}_{k\rightarrow \infty}$ is the same as the coefficient controlling the vacuum energy on $S^1 \times \mathbb{R}^2$. This is indeed the case in the examples considered. Interestingly, though, supersymmetric partition functions on $S^3$ (whose leading piece is not extensive) also have their high-temperature limit controlled by the vacuum energy on a lens space. 

\subsection{Further applications of large lensing in quantum field theory}
The primary assumption used throughout this work is an approximate equivalence between a nontrivial circle fibration and a trivial circle fibration. This can be used for applications unrelated to modular invariance. For example, we can consider CFT$_3$ on $S^3$. If the approximate equivalence is correct, then $Z[S^3/\mathbb{Z}_{p\rightarrow\infty}]\approx Z[S^1_{2\pi/p\rightarrow 0}\times S^2]$. Again, the equality should be between the leading singularity in the two path integrals. This limit gives credence to the idea of large-$N$ ``topological volume-independence" of center-symmetric gauge theories introduced in \cite{Shaghoulian:2016xbx} by relating it to ordinary large-$N$ volume independence \cite{Eguchi:1982nm, Kovtun:2003hr, Kovtun:2004bz, Kovtun:2007py}, which is firmly established.

Another potential application is to counting problems. We know that the finite, cutoff-independent part of $F = -\log |Z_{S^3}|$ \cite{Jafferis:2011zi} is a good monotonic quantity \cite{Casini:2012ei}. This free energy admits a counting interpretation in CFT since it can be conformally mapped to the entanglement entropy across an $S^1$ \cite{Dowker:2010yj, Casini:2011kv}. It is also known that the thermal coefficient $c_{\text{therm}}$ which enters the entropy density on $S^1 \times \mathbb{R}^{d-1}$ as $s \sim c_{\text{therm}} T^{d-1}$ (which is the same as the coefficient $\varepsilon_{\text{vac}}$ which enters into the vacuum energy on a spatial $S^1 \times \mathbb{R}^{d-2}$) is \emph{not} a good monotonic quantity unless $d=2$. This is a bit surprising since this is a good count of degrees of freedom if there ever was one. These two quantities can now be connected by the one-parameter family of lens spaces. The free energy on $S^3/\mathbb{Z}_p$ with thermal boundary conditions on the Hopf fiber as $p\rightarrow \infty$ becomes the thermal free energy on $S^1_{2\pi/p} \times S^2$. But this gives precisely $c_{\text{therm}}$ up to geometric factors. Understanding this one-parameter family may give intuition into the mechanics of the $F$-theorem and connect it to the breakdown of the $c_{\text{therm}}$-theorem. One conclusion we can immediately draw is that $F=-\log |Z_{S^3/\mathbb{Z}_p}|$, the lens space free energy, is not a good monotonic quantity for arbitrary $p$.

\section{Acknowledgments}
I would like to acknowledge Brian Willett for many helpful conversations and comments on a draft. I would also like to thank Arash Ardehali, Martin Fluder, Simeon Hellerman, Masazumi Honda, and David Morrison for useful conversations. This work is supported by NSF Grant PHY13-16748.

\footnotesize
\bibliographystyle{apsrev4-1long}
\bibliography{LensSpaceBIB}

\end{document}